\documentclass[useAMS,usenatbib]{mn2e}
\usepackage{psfig,morefloats,url,amsmath}
\usepackage{graphicx,subfloat}
\usepackage{subfig}

\footnotesize
\newdimen\digitwidth    
\setbox0=\hbox{\rm0}
\digitwidth=\wd0
\catcode`!=\active
\def!{\kern\digitwidth}

\normalsize

\title[FRBs at decametric wavelengths]{Detecting fast radio bursts at 
decametric wavelengths}

\author[Rajwade \& Lorimer]{
K. M. Rajwade$^{1,2}$\thanks{E-mail: kmrajwade@mix.wvu.edu} and
D. R. Lorimer$^{1,2,3}$\thanks{E-mail: duncan.lorimer@mail.wvu.edu}
\\
$^{1}$Department of Physics and Astronomy, West Virginia University, Morgantown,WV 26506, USA\\
$^{2}$Center for Gravitational Waves and Cosmology, West Virginia University, Chestnut Ridge Research Building, Morgantown, WV 26505, USA\\
$^{3}$Green Bank Observatory, Green Bank, WV 24944, USA\\
}

\pagerange{} \pubyear{2016}

\def\LaTeX{L\kern-.36em\raise.3ex\hbox{a}\kern-.15em
    T\kern-.1667em\lower.7ex\hbox{E}\kern-.125emX}

\begin{document}

\label{firstpage}

\maketitle

\begin{abstract}
Fast radio bursts (FRBs) are highly dispersed, sporadic radio pulses that are
likely extragalactic in nature. Here we investigate the
constraints on the source population from surveys carried out at
frequencies $<1$~GHz. All but one FRB has so far been discovered
in the 1--2~GHz band, but new and emerging
instruments look set to become valuable
probes of the FRB population at sub-GHz frequencies
in the near future. In this paper,
we consider the impacts of free-free absorption and multi-path
scattering in our analysis via a number of different assumptions 
about the intervening medium. We consider previous low frequency surveys alongwith an ongoing survey 
with the University of Technology digital backend for the Molonglo Observatory Synthesis Telescope (UTMOST) as well as future observations with the Canadian Hydrogen Intensity Mapping
Experiment (CHIME) and the Hydrogen Intensity and Real-Time Analysis
Experiment (HIRAX). We predict that CHIME and HIRAX will be able to observe
$\sim$ 30 or more FRBs per day, even in the most extreme scenarios
where free-free absorption and scattering can significantly impact
the fluxes below 1~GHz. We also show that UTMOST will
detect 1--2 FRBs per month of observations. For CHIME and HIRAX, the detection
rates also depend greatly on the assumed FRB distance scale.
Some of the models we
investigated predict an increase in the FRB flux as a function of redshift
at low frequencies. If FRBs are truly cosmological sources, this effect may impact 
future surveys in this band, particularly if the FRB population traces
the cosmic star formation rate.
\end{abstract}


\section{Introduction}

The origin of fast radio bursts (FRBs) remains an unanswered
question since their discovery a decade ago \citep{Lo07}. FRBs
are millisecond duration, highly sporadic and dispersed radio pulses
which follow the same dispersion relation seen in radio pulses from
neutron stars.  Of the 20 FRBs known so far, 18 have been found
at Parkes~\citep{Th13,Lo07,Pe15,Ke16,Ch16}, one at Arecibo
\citep{Sp14,Sp16} and one at Green Bank \citep{Ma15}. With the
exception of the latter, FRB~110523, which was detected at
800~MHz, all the other FRBs have so far been seen in the
1--2~GHz band.  FRB dispersion measures (DMs) are substantially
greater than that expected from free electrons in our Galaxy, suggesting
that FRBs are extragalactic in origin. There have been arguments about
local origin of FRBs but the models cannot explain all the observed
characteristics~\citep[for a review, see][]{Ka16}.

Broadly speaking, the FRB source models fall into two categories:
those of a catastrophic nature which would only be seen once~\citep[e.g.,
prompt emission from a gamma-ray burst;][]{Ya16} or those with the
possibility to repeat \citep[e.g., giant pulses from Crab-like
pulsars;][]{Co16,Co04}. So far, the only source known to repeat is
FRB~121102~\citep{Sp16}. In the light of these recent
discoveries, and to try to shed light on the origins of FRBs a number
of groups are carrying out extensive radio surveys at sub-GHz
frequencies~\citep{Ka15,Ca16,De16}. To date, however, the 0.7--0.9~GHz
detection of FRB~110523 remains the only source found below
1~GHz~\citep{Ma15}. 

\cite{Ly16} argues that a lack of detections could be due to absorption
in an ionized medium along the line of sight.  Recent discoveries
suggest low scattering in all FRBs which precludes a local plasma in
the vicinity of the progenitor to explain their high
DMs~\citep{Ma15,Ma13}.  \cite{Ku15} argue for a young magnetar model
with circum-dense medium around the star which can explain the high DM
and the non-detections at lower frequency due to free-free
absorption. The non-detections can also be explained by young neutron
star progenitor within an expanding supernova remnant shell with hot
ionized filaments~\citep{Pi16}.

In this paper, we present a detailed analysis of the aforementioned
absorption and scattering models. We use the approach to investigate
the significance of
non-detections in three recently completed surveys to constrain the
spectral index of the burst for each model. Based on these constraints
we make predictions for FRB detections from CHIME, UTMOST and HIRAX.~\cite{Co16b} make optimistic 
predictions for these upcoming low frequency surveys based on single FRB 
detection in the 0.7--0.9 GHz band. Here, we present predictions on 
the FRB detection rates based on different models of flux mitigation in the 
ISM. The plan for this rest of this paper is as follows. We describe our 
analysis methods in \S 2. In \S 3, we describe our results and discuss their 
implications in \S 4.

\section{Methods}

This section describes the methodology used for obtaining upper limits
on FRB predictions with CHIME under different astrophysical
scenarios. Our study is motivated by our recent work on modeling
gigahertz peaked spectrum pulsars via free-free
absorption~\citep{Ra16}. Here, we investigate what could happen to an
FRB that is absorbed or scattered and how that affects detectability
with CHIME and UTMOST. We will begin by making use of the recent null
results of FRB detections in the ongoing UTMOST survey~\citep{Ca16},
the Arecibo drift scan survey \citep[AO327;][]{De16} and the 145~MHz
LOFAR survey~\citep{Ka15}. We also considered the 155 and 182~MHz
surveys with the Murchison Widefield Array (MWA)~\citep{Ti15,Ro16} in our
analysis. However, since the flux limits for those surveys are higher than the LOFAR survey, the spectral index
constraints are less stringent than the LOFAR survey. We do not
include results from single-pulse searches in the ongoing Green Bank
North Celestial Cap (GBNCC) survey \citep{St14} in this analysis. A
paper describing the constraints from these results will be presented
elsewhere (Chawla et al., in prep). 

\subsection{Flux--redshift relationship and baseline model}
\label{sec:basicmodel}

Our methodology builds upon that used
by~\cite{Ka15} in their LOFAR survey, to include the effects of
free-free absorption and scattering. Following these authors,
we assume that FRBs are standard candles and the
energy output from the source follows a power law with respect to
frequency~\citep[see, e.g.,][]{Lo13}. Then, the peak flux density
\begin{equation}
S_{\rm peak} = \frac{L~\int_{\nu_{1}(1+z)}^{\nu_{2}(1+z)} E_{\nu'}
  d\nu'}{4\pi D(z)^{2}~\left(\nu_{2} - \nu_{1}\right)~\int_{\nu'_{\rm low}}^{\nu'_{\rm high}} E_{\nu'}
  d\nu'},
\label{eq:speak}
\end{equation}
where $L$ is the bolometric luminosity, the pulse energy
$E_{\nu'} \propto \nu'^{\alpha}$ for some
spectral index $\alpha$ and source frame frequency $\nu'=(1+z)\nu$ at
redshift $z$ and luminosity distance $D(z)$. Also in the source frame,
$\nu'_{\rm low}$ and $\nu'_{\rm high}$ are the frequency bounds in which
the source emits. Following~\cite{Lo13},
we assumed $\nu'_{\rm high}=$10~GHz and $\nu'_{\rm low}=$10~MHz.
The observed frequency band is defined by $\nu_1$ and $\nu_2$ and is
different for each survey under consideration. We will discuss the 
implications of this standard-candle assumption in \S \ref{sec:caveats}.

\begin{figure}
\centering
\centerline{\psfig{file=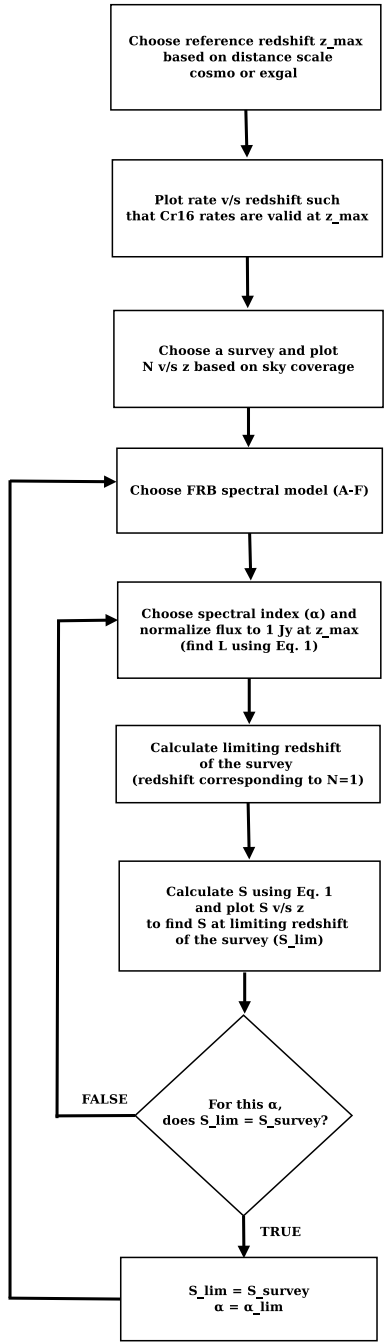,width=5cm}}
\caption{Flow diagram showing the logical flow of our analysis procedure
for placing constraints on the spectral index. For further details, see
\S \ref{sec:basicmodel}.}
\label{fig:frbflow}
\end{figure}

Our implementation of the earlier study by \cite{Ka15} to 
place constraints on FRB spectral indices is summarized 
in Fig.~\ref{fig:frbflow} and described below. Since the
distance scale for FRBs is not well known, we consider two scenarios:
(i) a ``cosmological model'' for which the maximum redshift
$z_{\rm max} = 0.75$~\citep[see, e.g.,][]{Lo13}; (ii) an
``extragalactic model'' for which the characteristic distance is
100~Mpc (i.e.~$z_{\rm max} = 0.025$; see, e.g. Lyutikov et
al.~2016). Having chosen one of these two scales, we then
derive the FRB rate versus redshift relationship by assuming
an FRB population with constant density per unit comoving
volume out to $z_{\rm max}$. At the chosen value of $z_{\rm max}$
this rate matches, by definition, the rates 
published by~\cite{Cr16} based on FRB surveys at Parkes. 
Using this curve, for each of the other surveys under consideration
(i.e.~LOFAR, AO327 and UTMOST), we can compute the number of FRBs
expected as a function of redshift by multiplying the rate--redshift
relationship by the appropriate survey sky and time coverage. The
resulting number versus redshift curves then lead to a limiting
redshift $z_{\rm lim}$ for each survey. This limiting redshift is
defined to be that at which $<1$ FRB is predicted to be seen in 
each survey. An example of one such calculation
is shown for UTMOST in the left panel of Fig.~\ref{fig:speak}.

Next, for each of the
source models A--F described in detail below, we choose a
spectral index $\alpha$ and, using Eq.~1, find the corresponding
value of $L$ such that $S_{\rm peak}=1$~Jy at $z_{\rm max}$.
The 1~Jy reference flux is approximate, and motivated by the results
of Thornton et al.~(2013). Our results turn out to be insensitive
to the exact value adopted here.
For each of the surveys under consideration, we calculate
the corresponding flux at the survey's redshift limit, 
i.e.~$S_{\rm peak}(z_{\rm lim})$ and iterate until the spectral
index is found where $S_{\rm peak}(z_{\rm lim})$ equals
the survey flux limit. This spectral index is, by definition,
the limiting value appropriate to the assumptions of that
particular model and distance scale, and we refer to this 
lower limit as $\alpha_{\rm lim}$.

Our baseline model, which follows this process using a simple
power-law spectral behaviour amounts to a repeat of the analysis
of~\cite{Ka15}. We refer to this case as model ``A''
henceforth and, as necessary, distinguish between the cosmological
and extragalactic cases in the text. The relevant parameters used for each of these models and
constraints obtained from them are given in
Table~\ref{tab:params} and discussed further in the sections below.

\subsection{FRB survey sensitivity model}

From radiometer noise considerations, if $W$
is the width of the FRB then, for a search in which the
pulse is optimally match filtered by a top-hat pulse of peak
flux density $S$, the signal-to-noise ratio
\begin{equation}
{\rm S/N} = \frac{S \, G \sqrt{W n_{\rm p}\Delta\nu}}{T_{\rm sys}},
\label{eq:flux}
\end{equation}
where $T_{\rm sys}$ is the system temperature, $\Delta\nu$ is the
bandwidth, $n_{\rm p}$ is the number of polarizations summed and $G$
is the gain. In all current FRB surveys, where incoherent dedispersion
techniques are used to process the data, and in the context of our
models DM depends on redshift, then there is a dispersive broadening
effect that results in a dependence between survey sensitivity and
redshift. To model this effect, we compute the effective width of the pulse
\begin{equation}
W_{\rm eff} = \sqrt{W_{\rm int}^{2} + W_{\rm DM}^{2} + W_{\tau}^{2}},
\end{equation}
where $W_{\rm int}$ is the intrinsic pulse width of the FRB, $W_{\rm DM}$ is
the intra-channel dispersion smear and $W_{\tau}$ is the additional
broadening due to the finite sampling interval of the survey. To calculate
$W_{\rm DM}$, we adopted a DM-redshift scaling from~\citep{In04} where
DM~$= 1200~z$~cm$^{-3}$~pc. Using the standard expression for
dispersion broadening \citep[see, e.g.,][]{Lo04}, we have
\begin{equation}
  W_{\rm DM} = 99.6 \, {\rm ms} \left(\frac{z}{n_{\rm chan}}\right)
\left(\frac{\Delta \nu}{\rm MHz}\right)
\left(\frac{\nu}{\rm GHz}\right)^{-3},
\end{equation}
where $n_{\rm chan}$ is the number of frequency channels used
for dedispersion. Future FRB surveys may well introduce
high-speed algorithms to implement coherent dedispersion
\citep[see, e.g.][]{Za14}, in which case $W_{\rm DM}$ will not
be necessary. To model the degradation due to incoherent dedispersion
of current and near-future surveys, consider an ``optimal survey'' signal-to-noise
ratio, S/N$_0$ which is obtained from Eq.~\ref{eq:flux} for the case for
a top-hat pulse with height $S_0$ and width $W_{\rm int}$. For
a broadened pulse of width $W_{\rm eff}$, energy conservation means
that its peak flux density is $S_0 W_{\rm int}/W_{\rm eff}$. It is
straightforward to show that the S/N
of the broadened pulse is lower than S/N$_0$
by a factor of $\sqrt{W_{\rm int}/W_{\rm eff}}$. For an actual
survey with a constant S/N threshold, this amounts to an {\it increase}
in the limiting peak flux density for detection by the reciprocal
of this factor, so that the resulting limiting sensitivity
\begin{equation}
S_{\rm lim} = S_0 \sqrt{\frac{W_{\rm eff}}{{W_{\rm int}}}} = 
\frac{{\rm S/N}_{\rm lim} \, T_{\rm sys}}{G \, W_{\rm int}}\sqrt{\frac{W_{\rm eff}}{n_p \Delta \nu}}.
\end{equation}
This expression is used when calculating the sensitivity curves
throughout this paper (see, e.g., the right panel of Fig.~\ref{fig:speak}).
Here S/N$_{\rm lim}$ is the limiting signal-to-noise ratio 
required for a detection in a given survey. 
Table~\ref{tab:FRBsurv} summarizes the essential
observing parameters for each of the surveys considered in this paper.

\begin{table*}
\begin{tabular}{lc lc lc lc lc}
\hline
Survey & Centre frequency & Bandwidth & Flux limit & Reference   \\
      & (MHz) & (MHz) & mJy &  \\
\hline
UTMOST& 843& 31.5 &11000 & ~\cite{Ca16}\\
AO327& 327& 57 & 83  & ~\cite{De16}\\
LOFAR & 145 & 6 & 62000 & ~\cite{Ka15}\\
CHIME & 600 &400 &125 & ~\cite{Ne14}\\
HIRAX&600 & 400& 24 & ~\cite{Ne16} \\
\hline
\end{tabular}
\caption{Table showing various parameters of different surveys. The system parameters of CHIME and HIRAX are estimated values (see text for details).
\label{tab:FRBsurv}}
\end{table*}

\section{Models for flux mitigation}
Radio signals propagating through the ISM are modulated by free 
electrons in the intervening medium. These interactions leave observational
signatures in the received radiation at the earth. Some of these signatures
(e.g.~scattering, free-free absorption and scintillation) have been observed in 
various radio sources. FRBs, being astrophysical in nature, are subject
to the same phenomena. It is therefore important to model these effects in 
detail before we draw any inferences about their intrinsic spectral indices and 
make predictions for future surveys. Below, we describe our mathematical
models to characterize effects of scattering and free-free absorption.

\subsection{Models including free-free absorption}

As discussed by other authors~\citep{Ku15,Ly16}, but not taken into
account by \cite{Ka15}, thermal absorption can significantly
reduce FRB fluxes at lower frequencies. For this analysis, following
our earlier work~\citep{Ra16}, we assume
\begin{equation}
E_{\nu'} \propto \nu'^{\alpha}~{\rm exp}\left(-\tau \nu'^{-2.1}\right),
\label{eq:E}
\end{equation}
where, as described further by \cite{Ra16}, the optical depth of the absorber
\begin{equation}
\tau = 0.082~T_{e}^{-1.35}~{\rm EM}.
\end{equation}
Here $T_{e}$ is the electron temperature and EM is the emission
measure of the absorber. Then the peak flux is computed by combining
Eq.~\ref{eq:speak} and Eq.~\ref{eq:E}. We consider two cases for
absorption: (i) cold, molecular clouds with ionization fronts for which
$T_e = 200$~K and EM~$=1000$~cm$^{-6}$~\citep{Le15} (hereafter, 
model B); (ii) hot, ionized magnetar ejecta/circum-burst medium 
for which $T_e= 8000$~K and EM~$=1.5 \times 10^6$~cm$^{-6}$
(hereafter, model C). The value of EM for model C has been chosen from a range of values 
reported in~\cite{Ra16}, \cite{Ku15} and \cite{Le15}.

\begin{table*}
\begin{tabular}{lc lc lc lc lc lc lc}
\hline
Model & $T_{e}$ & EM & & \multicolumn{3}{c}{$\alpha_{\rm lim}$} & & & \multicolumn{2}{c}{$z_{\rm lim}$}  \\
   & (K) & cm$^{-6}$pc & \multicolumn{2}{c}{UTMOST} & \multicolumn{2}{c}{LOFAR} & \multicolumn{2}{c}{AO327} & \multicolumn{2}{c}{CHIME} \\
\hline
  &  & & cosmo & exgal & cosmo & exgal & cosmo & exgal & cosmo&exgal \\
A & --- & ---            &--0.70 &--1.30&0.0& --0.50  & \,\,\,0.70 & \,\,\,1.25 & 1.54 & 0.10\\
B & 200& 1000            &--0.80&--1.30&--1.0&--2.10 &\,\,\,0.50&\,\,\,1.10 & 1.56& 0.10 \\
C &8000&$1.5 \times 10^6$&--1.50 &--2.50& --- & ---& --0.30 &--2.85 & 1.64 & 0.09 \\
D & --- & --- & --2.10 & --3.30 &--3.0& --4.0 &--3.30& --2.20 &0.84& 0.06 \\
E & 200 & 1000 & --2.20 &--3.30& --4.10 & --5.70 &--3.50& --2.50 & 0.85 & 0.06  \\
F & 8000 &1.5$\times$10$^{6}$&--2.70& --4.50& --- &---&--4.50&--6.45& 0.82& 0.05  \\
\hline
\end{tabular}
\caption{Model parameters and resulting spectral constraints from the
  various surveys considered. From left to right, we list the model,
  assumed electron temperature ($T_e$) and emission measure (EM),
  limiting spectral index ($\alpha_{\rm lim}$) for the three published
  surveys (LOFAR, AO327 and UTMOST). For the future CHIME survey, we
  list the limiting redshift ($z_{\rm lim}$) predicted by our models.
  The ``cosmo'' and ``exgal'' columns give results from the two
  different distance scales assumed: ``cosmological'' ($z=0.75$) and
  ``extragalactic'' ($z=0.025$) as defined further in the text.}
\label{tab:params}
\end{table*}

\begin{figure*}
\centering
\centerline{\psfig{file=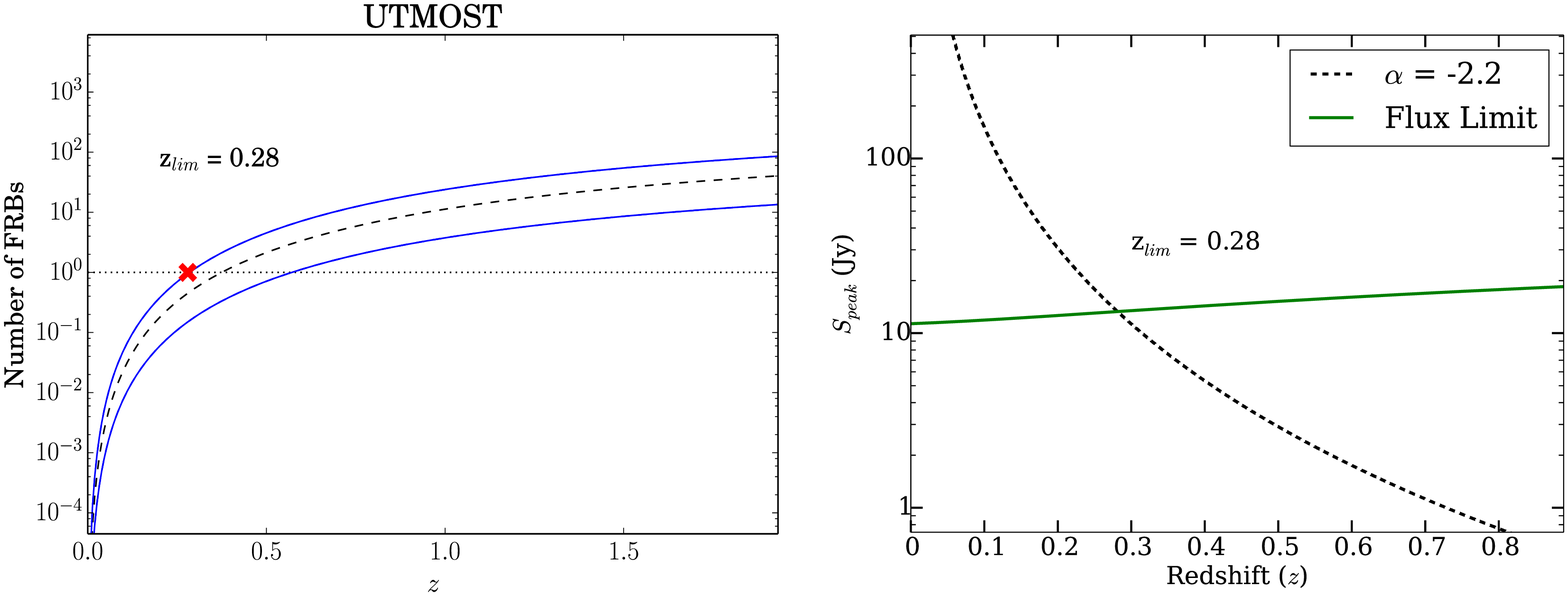,width=15cm}}
\caption{Left: Number of FRBs versus redshift for the UTMOST survey parameters~\protect\cite{Ca16}. The curves indicate
  the~\protect\cite{Cr16} rates with 99$\%$ bounds. The horizontal
  line corresponds to one FRB. The intersection of the horizontal line and 
the upper bound of the curve is
  shown by the red cross at $z = 0.28$. Right: peak flux versus
  redshift for UTMOST survey for model E (dashed curve). The solid
  line shows the flux limit of the UTMOST survey. The intersection of the two curves is denoted by z$_{lim}$. Note
  the non-linear dependence of flux limit with redshift for both
  surveys shown here is due to the impact on intra-channel dispersion
  broadening upon sensitivity (see \S 2.2 for details).}
\label{fig:speak}
\end{figure*}

\subsection{Models including multi-path scattering}

Multi-path scattering due to free electrons in the ionized medium
along the line of sight to the observer can cause a reduction in the
measured flux at the telescope. Scattering manifests itself as an
exponential tail in the radio pulse of the FRB. FRBs that have been
discovered so far, show only a modest amount of scattering: for the 17
FRBs, 10 of them have scattering measurements and 7 have them
have upper limits~\citep{Cor16}.  Hence, we computed the scattering
timescale by taking the average of the published values (estimates and
upper limits) of these 17 sources. For sources with upper limits, conservatively,
we assumed those values as measured values when taking the average. We obtained a mean 
scattering timescale of $\sim$8.1~ms at 1~GHz. We note that if we assume the 
scattering timescales for sources with upper limits as half of the upper limit 
values, we get a average timescale of $\sim$6.7~ms which is also a high value. Using the most conservative value, the scattering
timescale $\tau_{s}$ can be computed for any frequency $\nu$ via the
$\nu^{-4}$ scaling law~\citep{Bh04} as opposed to $\nu^{-4.4}$. The non-Kolmogorov scaling
exponent is due to fact that the diffraction length scale is smaller
than the inner scale of the wavenumber spectrum~\citep[see][, and references
  therein]{Bh04}. Assuming that energy of the burst is conserved,
if the pulse scatters with a timescale of $\tau_{s}$, the width
increases and hence, the measured flux reduces by a factor of
$\sqrt{1+(\tau_s/W_{\rm eff})^2}$ where $W_{\rm eff}$ is the effective
pulse width defined in the preceding section.
Including this effect into our analysis, we introduce three final
models. Model D has scattering with no free-free absorption, while
models E and F have scattering in addition to the respective
absorption parameters adopted for models B and C.

\section{Results}

\subsection{Spectral index constraints}

Taking into account all the factors discussed in the previous section,
the results of our analysis are collected for models A--F in 
in Table \ref{tab:params}.  For each of these models, we
constrained the spectral indices assuming each of the two 
distance constraints in turn. A graphical illustration of this
process is shown for model E as an example in 
Fig.~\ref{fig:speak} where we show the constrained spectral index for
one of the models for the UTMOST survey~\citep{Ca16}. As mentioned
previously, our baseline model (A) is an update on the 
results of \cite{Ka15} using the more recent rate FRB rate estimates
from \cite{Cr16}. In our analysis, which also includes the non-detections
in UTMOST and AO327, the most constraining power-law spectral index for
this model is $\alpha > 0.7$ for the cosmological distance scale from AO327.
The most constraining spectral index ($\alpha > 1.25$) is obtained from the 
AO327 survey if the extragalactic distance scale is applied to this survey.

In model B, where we go beyond the simple power-law spectral 
dependence and include free-free absorption with cold molecular
clouds, we find only a modest change in the results for model A
for AO327 and UTMOST but as expected a greater deviation at the LOFAR
frequency band where spectral turnover effects are more severe. 
The LOFAR survey does not in fact provide any constraints on the spectral
index for models C and F, where a hot ionized medium is assumed.
These models predict flux densities below the survey threshold 
for essentially all values of $\alpha>-10$.
The corresponding $\alpha_{\rm lim}$ values are therefore not listed
in Table 2. 

The spectral index constraints become much weaker when the effects of
interstellar scattering are incorporated in models D, E and F. For model
D, with scattering but no free-free absorption is assumed, the UTMOST
null results only bound $\alpha >-2.2$ for the cosmological case and the AO327 results bound 
$\alpha > -2.2$ for the extragalactic case. When free-free absorption 
and scattering are considered in models E and F, these constraints are
diminished further.

\subsection{FRB rate predictions for future surveys}

\begin{figure*}
\centering
\begin{tabular}{|@{}l|@{}l|r@{}|}
{\mbox{\psfig{file=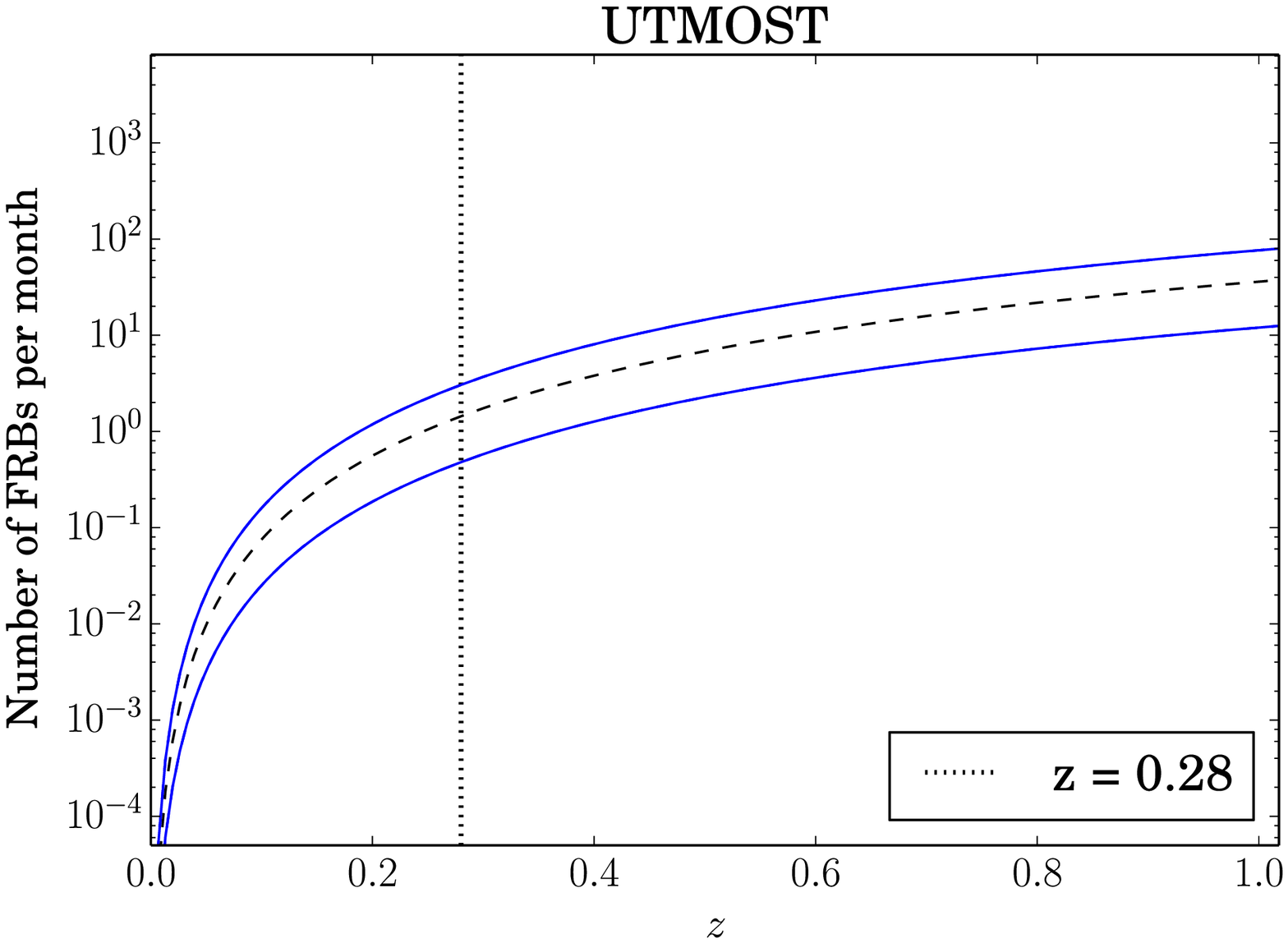,width=8cm}}} & {\mbox{\psfig{file=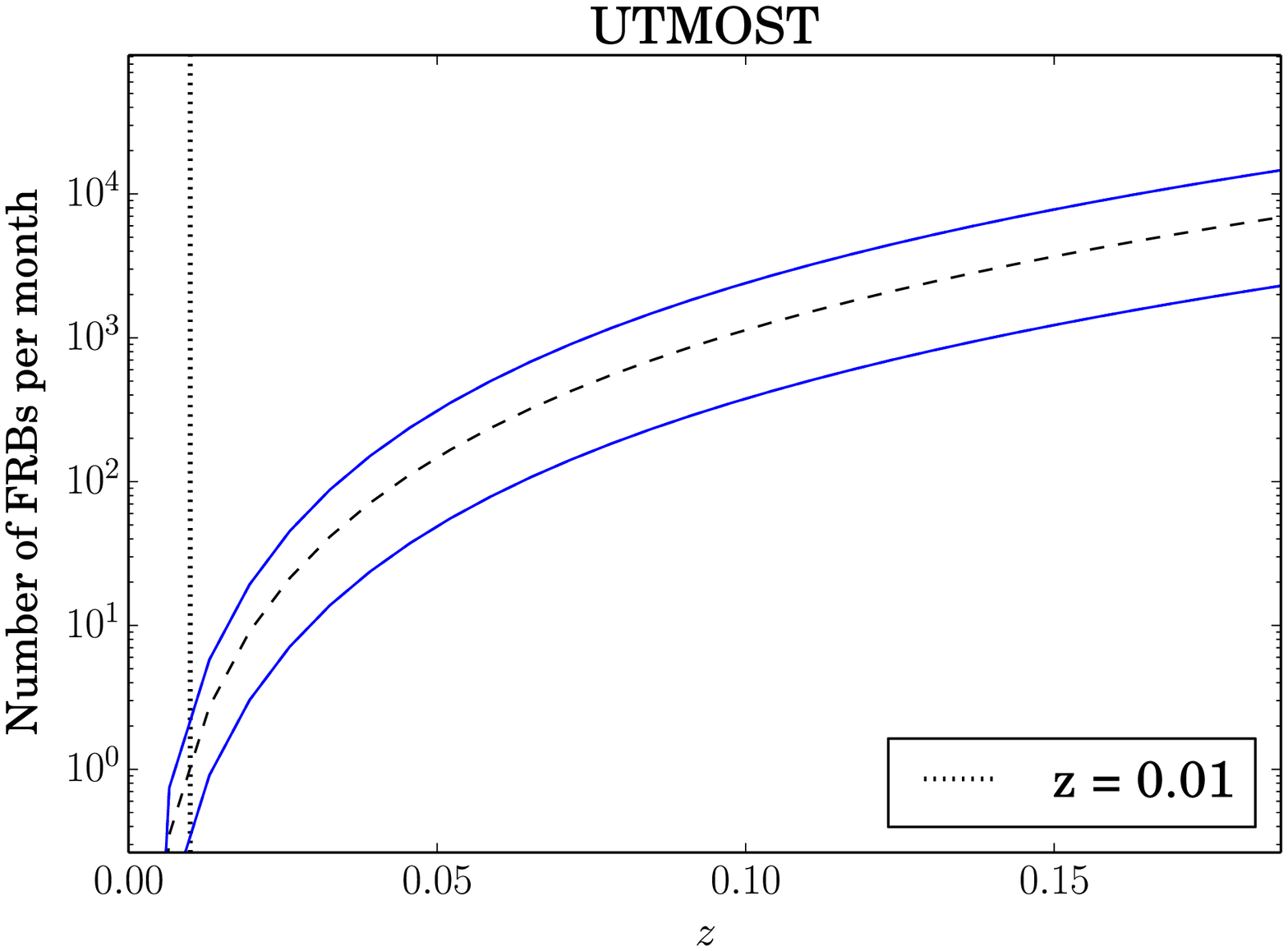,width=8cm}}}\\

{\mbox{\psfig{file=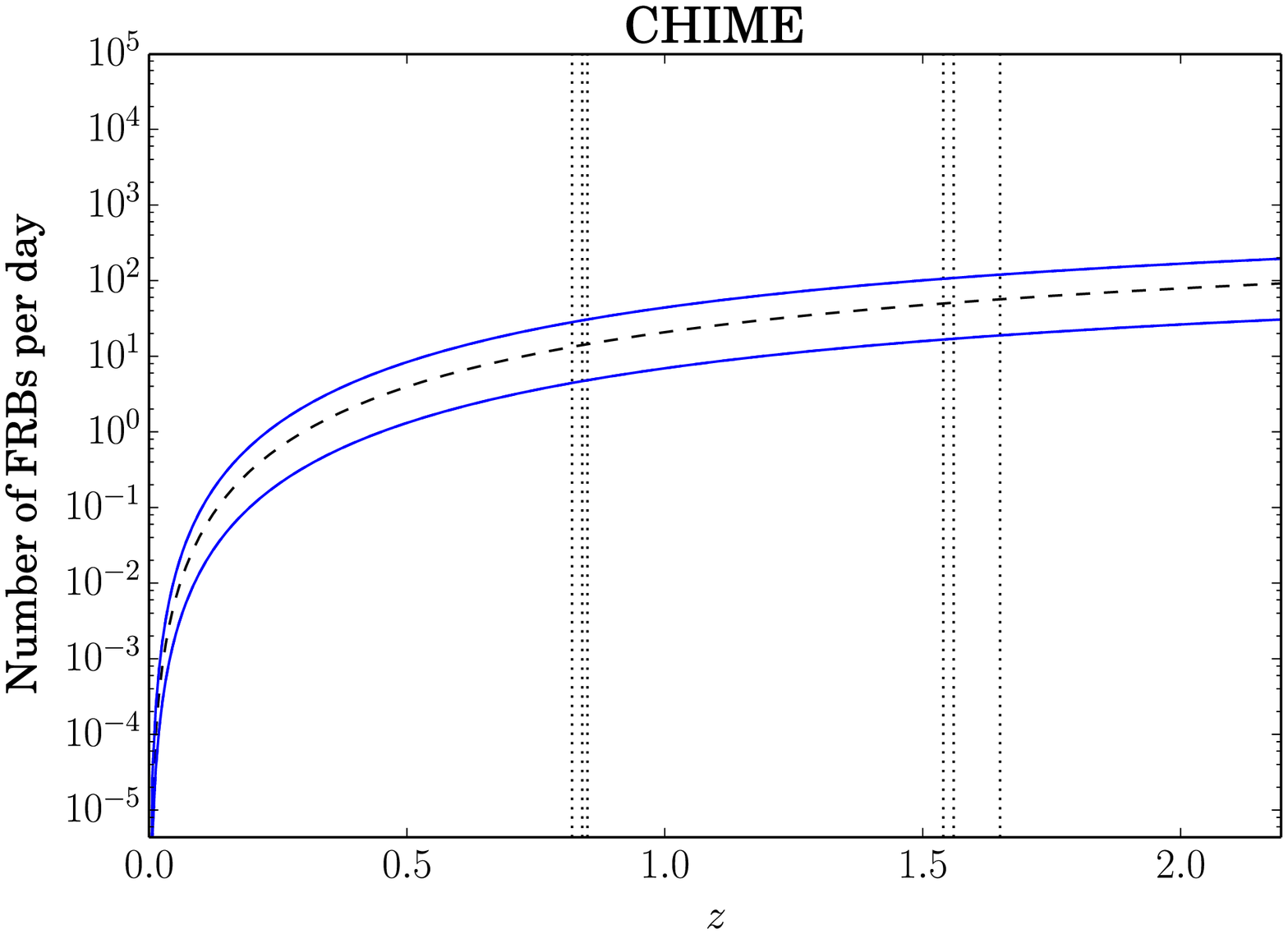,width=8cm}}} & {\mbox{\psfig{file=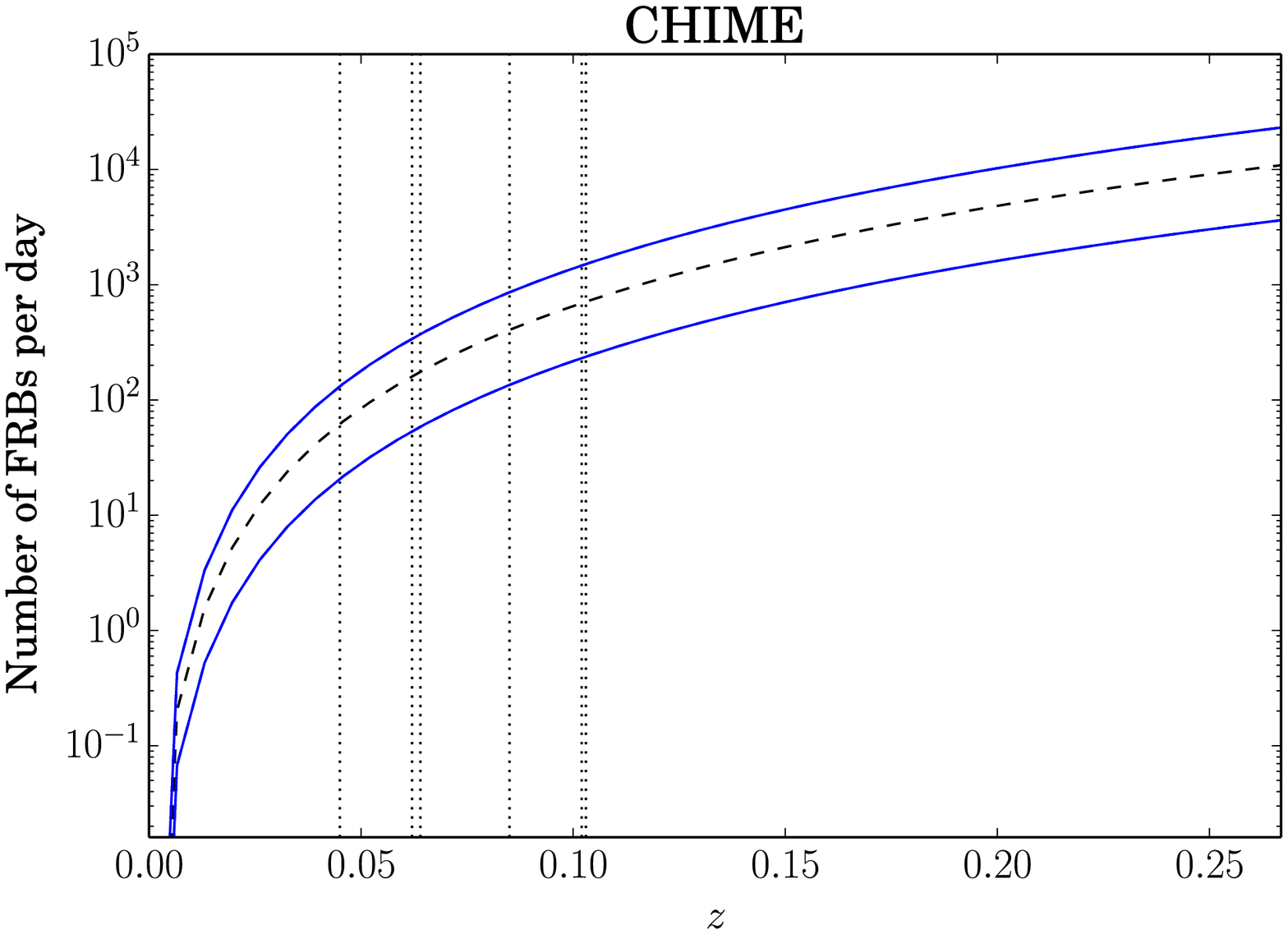,width=8cm}}}\\

{\mbox{\psfig{file=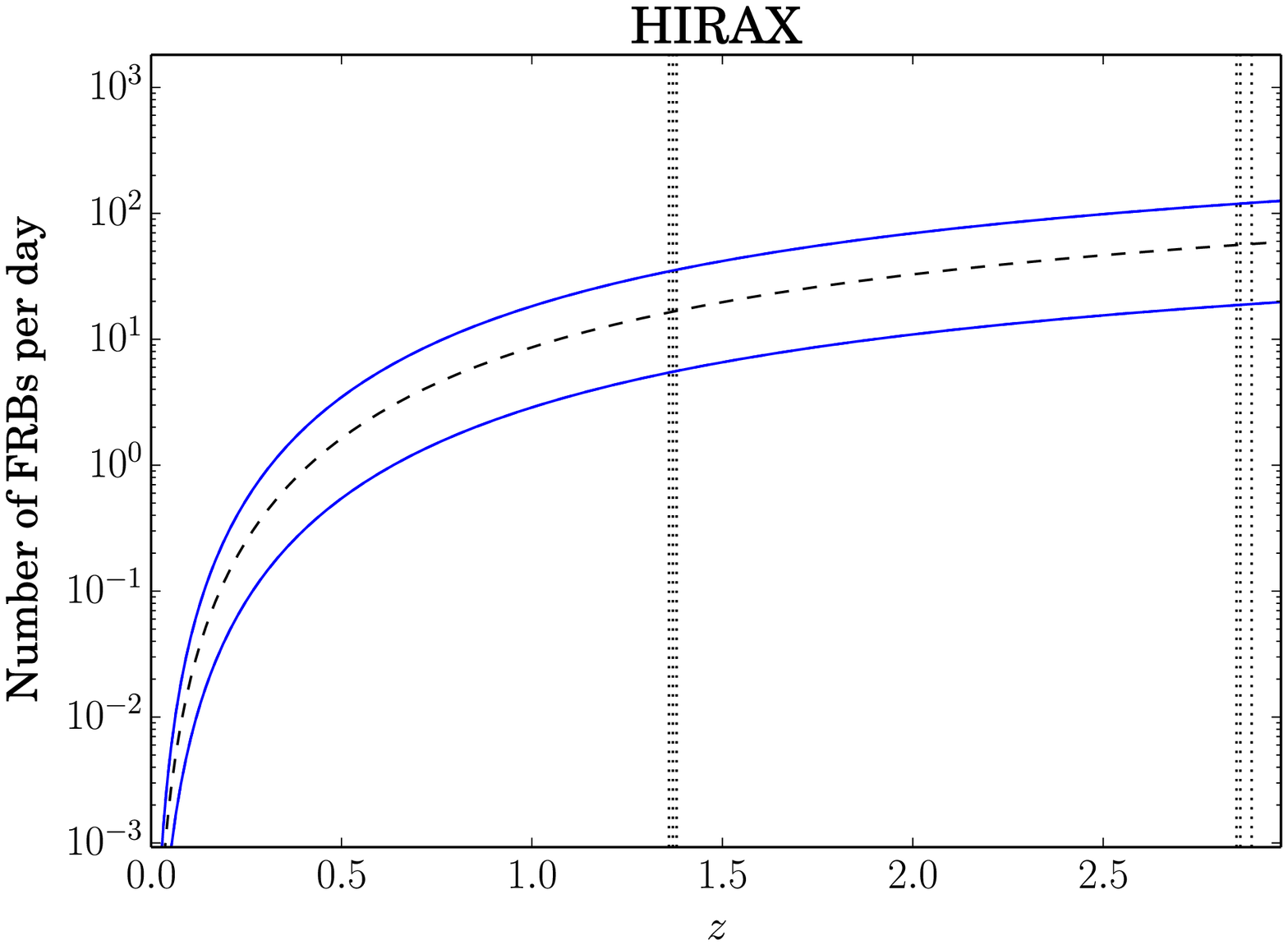,width=8cm}}} & {\mbox{\psfig{file=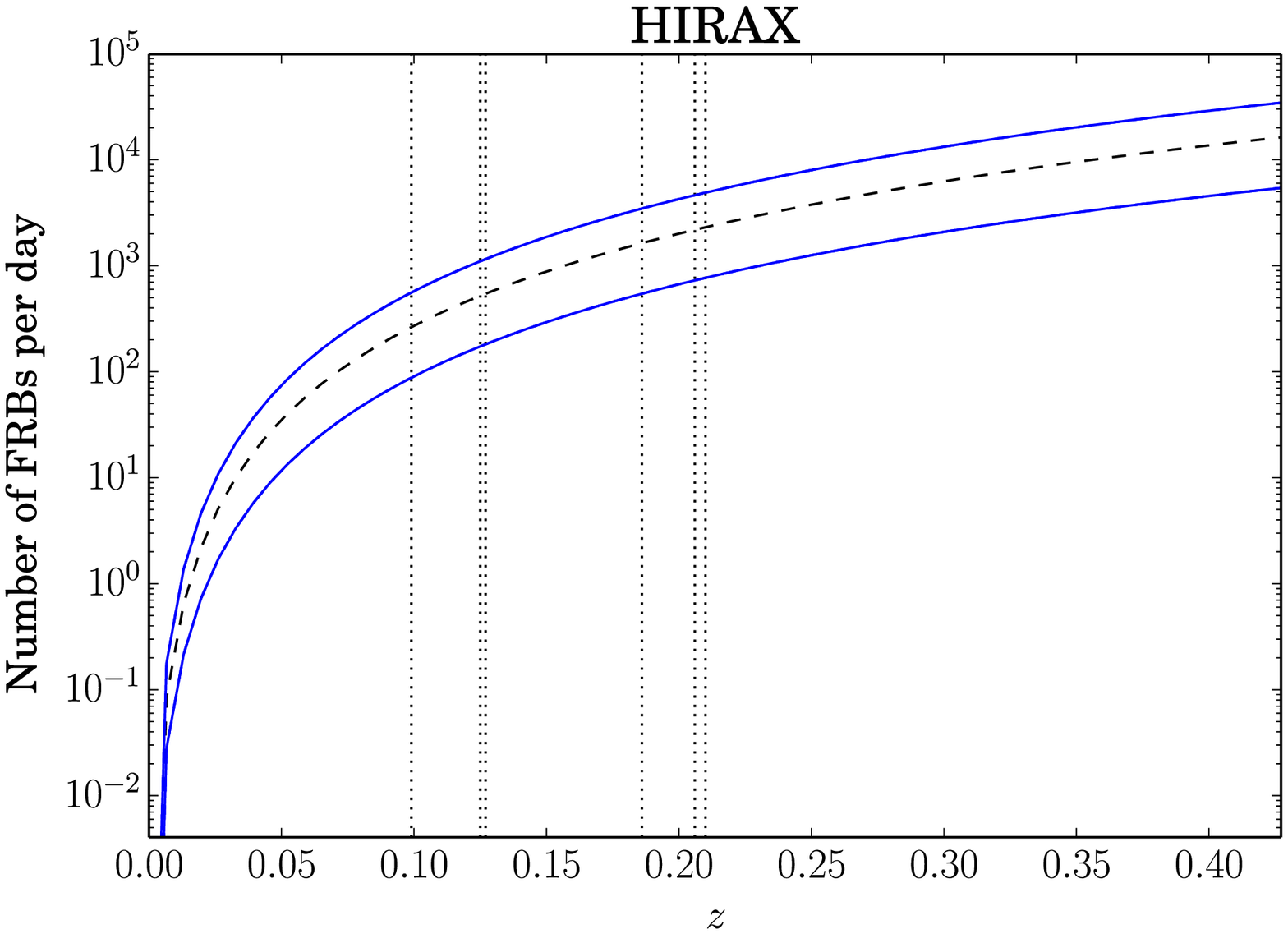,width=8cm}}}\\

\end{tabular}

\caption{The number of FRBs predicted per day/month as a function of
  redshift for various surveys. The black dashed curve is the number
  of FRBs per day based on the ~\citep{Cr16} rates. The blue curves
  are the 99$\%$ upper and lower bounds on the black dashed curve. Left panels show predictions for the cosmological case while
  the right panels show predictions for extragalactic case.  In
  predictions for CHIME (cosmological case), from left to right, the
  vertical lines correspond to models F,D,E,A,B and C respectively
  while they correspond to models F,E,D,C,B and A respectively for the
  extragalactic case. Similarly for HIRAX, the vertical lines from
  left to right correspond to models D,E,F,A,B and C respectively for
  the cosmological case and F,E,D,C,B and A respectively for the
  extragalactic case. In case of UTMOST, the single vertical line
  corresponds to all models for their respective constrained spectral
  index at the limiting redshift of the survey. The ordinate of
  intersection of the vertical line and the curves gives the predicted number 
for each  model. }
\label{fig:N_z}
\end{figure*}

Fig~\ref{fig:N_z} shows the
predicted detection rates for UTMOST, CHIME and HIRAX for the
two distance scales considered. The vertical
line corresponds to the redshift limit of the survey for all models
A--F. These predictions were obtained from the spectral constraints
on each model obtained in the previous section, and computing
the sensitivity of each survey as described below.

\begin{figure}
\centering
{\mbox{\psfig{file=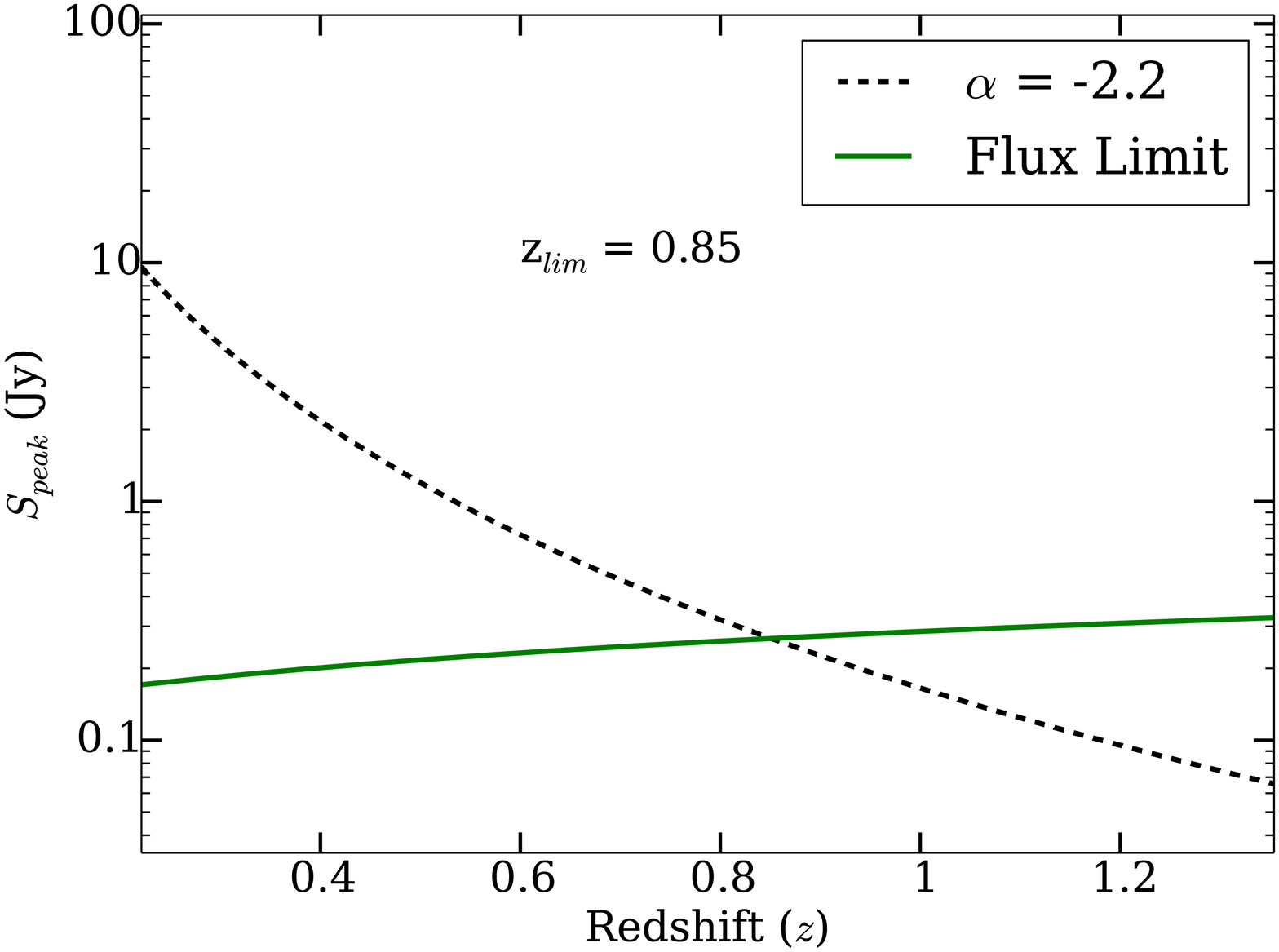,width=8cm}}}
\caption{Peak flux versus redshift predicted for model E
  assuming the nominal parameters of
  CHIME. The intersection of the survey limit and the curve gives the limiting redshift probed
by CHIME for this model. }
\label{fig:Speak_CHIME}
\end{figure}

In modeling the
sensitivity of CHIME, we assume that the gain $G=2$~K~Jy$^{-1}$ and
system temperature $T_{\rm sys}=50$~K remain constant over the band. We also
assumed a single CHIME beam of width 1.5 by 90 degrees (Bandura, private
communication).  Using Eq~\ref{eq:flux}, we obtained the optimum flux
limit of 0.125 Jy for a 5~ms duration burst. For the scattering
scenario, we used the frequency weighted average value of $\tau_{s}$
over the whole CHIME band. We obtained $\tau_{s}=$~92.2~ms.  For each
of the models described in Table.~\ref{tab:params}, and the using the
constraint on the spectral index from the UTMOST survey, we plotted
the peak flux versus redshift using Eq.~\ref{eq:flux}. For each model
at the constrained spectral index, we obtained the $z_{\rm lim}$ which
is the redshift where the peak flux of the FRB is equal to the flux
sensitivity limit of CHIME as shown in
Fig.~\ref{fig:Speak_CHIME}. Then, using the expected sky coverage of CHIME and
scaling the~\cite{Cr16} rate with the comoving volume, we obtained the
predicted number of FRB detections per day versus redshift as shown in
left panel of Fig.~\ref{fig:N_z}. The ordinate of the point at which
the $z_{\rm lim}$ for each model intersects the curve and the bounds
gives the predicted number of FRB detections per day for that given
model. We investigated the yield for HIRAX surveys with identical 
parameters as the ones used for CHIME except for $G$ = 10.5~K~Jy$^{-1}$. 
The analysis suggests that CHIME will be able to detect from 30--100
FRBs per day depending on the model for the cosmological case while
the yield increases by an order of magnitude ($\sim$150--1000 FRBs per
day) for the extragalactic case due to the sharp dependence of rates
with redshift. Similarly, HIRAX will be able to detect 50--100 FRBs per
day for the cosmological case and 700--4000 FRBs per day for the
extragalactic case.

\subsection{Caveats}
\label{sec:caveats}

Our analysis has a number of simplifying assumptions about the nature
of FRBs. In this section, we investigate the sensitivity of our
results to these assumptions. A key simplification we have made is to
assume that FRBs are standard candles. Recent models and surveys for
FRBs suggest that there might be distribution of luminosities for
these bursts~\citep[see, e.g.,][]{Ca16,Ve16}. Hence, we investigated the 
effect of FRBs having a range of luminosities. By definition, for a
population of standard candles, all sources are detected out to a
survey's redshift limit. This means that, for a distribution of luminosities,
only those FRBs that are fainter than the currently assumed value
will have any impact on the results. To investigate this, we repeated
our analysis by reducing the luminosities by a factor of 10 from
the value assumed above. This factor is motivated by the approximate
distribution of energies in the study of \cite{Ca16}. This exercise
resulted in weaker constraints on the spectral index values for each model
such that the $\alpha_{\rm lim}$ values reported in Table 2 are reduced
by factor of anywhere between 1.5 and 2 . Therefore, for a population with a
range of luminosities in general, we would expect the constraints 
given in Table 2 to be reduced slightly. We also note that lowering
the luminosities assumed necessarily results in lower predicted yields
for future FRB surveys. For example, we found that our predictions for
CHIME were reduced by up to a factor of 2. In summary, a range of luminosities
for the FRB population will tend to reduce the constraints on spectral
index and lead to different survey yields. This complication only further
highlights the value that future surveys will have in probing the FRB
population.

The recent discovery of a repeating FRB \citep{Sp16} provides some
evidence that a neutron star scenario is the most plausible model for these
bursts. If FRBs do originate from neutron stars, we detect the
brightest pulses from them in the local universe. This
constrains the distance to these sources to $z = 0.025$ (i.e.~100 Mpc). 
We also investigated the effect of such an assumption and
results are shown in Table~\ref{tab:params} and Fig.~\ref{fig:N_z}.
One would assume that given a smaller distance to the sources, CHIME
would see more of them. The results agree with this conjecture.
Fig.~\ref{fig:N_z} suggests that even with models including scattering
and free-free absorption,
CHIME would see $\sim$100 FRBs per day if they were in the local
universe. 

In all of our calculations, we have implicitly assumed that the
FRB rate is constant per unit comoving moving probed by the surveys.
If the FRB rate traces the cosmological star formation rate (SFR), then we
would expect the maximum number of sources to be found at $z=2$ 
(5.3 Gpc)~\citep{Ma14}.~\cite{Ca16b} compared a sample population of FRBs 
based on the SFR to the observed sample and found a good match with 
different parameters of the observed sample although the pulse widths 
could not be accounted for. Given the current size of the FRB population,
and difficulties in ascribing a distance scale, we regard this
as a subtlety that is currently not well probed by the observations.
We do, however, comment on a related factor that may impact future
observations in the discussion below.

\begin{figure*}
\centering
{\mbox{\psfig{file=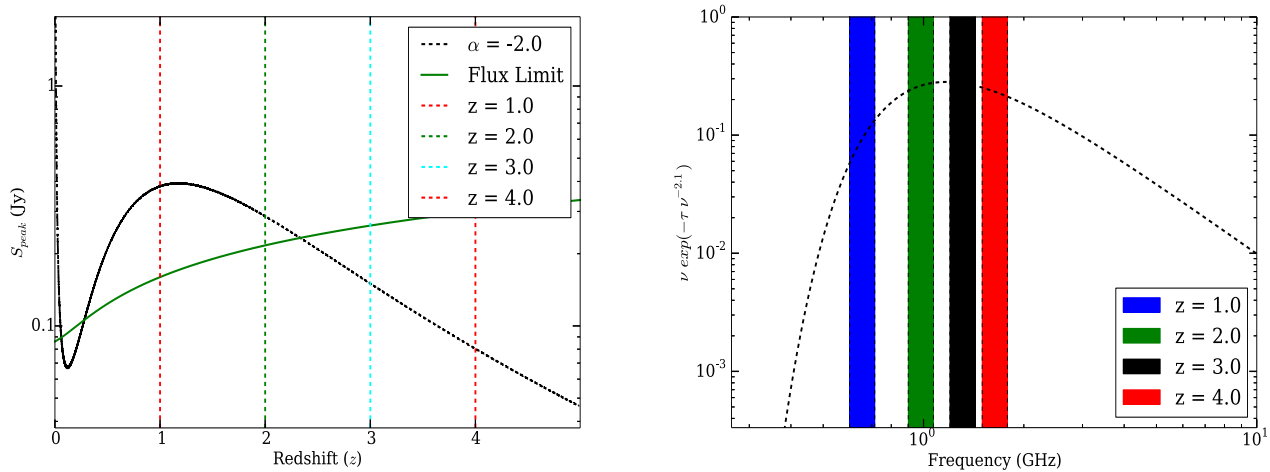,width=18cm}}}
\caption{Left: peak flux versus redshift for the AO327 survey for model F illustrating
  the effect of absorption and Doppler shift of the observed frequency as
  described in the text. The black dashed line is the flux of the FRB. The different vertical lines correspond to different redshifts.  In
  this case, we assumed $\alpha = -2$, EM =
  3$\times10^{6}$~cm$^{-6}$~pc and $T_{e} = 8000$~K. Right: the
  different regions of the absorption spectrum probed by the survey at different
  redshifts. The different shaded regions correspond to the rest frame frequency probed by the survey at different redshifts. }
\label{fig:spec}
\end{figure*}

\section{Discussion}

Our results suggest that telescopes in the 0.4--1.0~GHz band will make
vital contributions to our understanding of FRBs. Even with free-free
absorption and scattering playing a vital role in flux mitigation of
FRBs, CHIME will be able to detect these bursts on a
daily basis by the virtue of its extensive bandwidth and vast
instantaneous sky coverage. We also looked into the possible caveats
in the analysis and the effects those would have on the predictions
for CHIME. Our investigation suggests that with all the caveats
considered, the lowest yield for a future CHIME survey is $\sim 30$
FRBs per day which is very optimistic compared to expected yield from
other surveys. For example, the corresponding yield for future UTMOST
observations is about 1--2 FRBs per month for future observations
which makes it difficult to differentiate between the two models at the
moment.

We also discussed certain caveats in our analysis (\S \ref{sec:caveats})
and how these assumptions affect the results. We found that a distribution
in luminosities for FRBs, rather than a standard candle model assumed here,
results in weaker constraints for the spectral indices of the population.
Future surveys, however, will be excellent at probing the FRB luminosities
through the dependence of luminosity on survey yield.

If the FRBs currently observed
lie predominantly in the local Universe (i.e.~have characteristic 
distances of 100~Mpc), then the large DMs 
cannot be accounted for by the Milky Way, host and IGM
contributions. This discrepancy suggests that a large contribution to
the DM comes from the local plasma around the source which favours
models C and F as the most plausible scenarios
describing these events. Assuming the parameters in model C, we
can estimate the linear size of the absorber around the
source in order to produce the high DMs observed for FRBs.
If we take the FRB with the highest known DM (FRB 121002) and
place it at $z = 0.025$ then, assuming model C, we obtain a linear
size of $\sim$ 1.4~pc. This is very similar to the
parsec size high density filaments found in supernova remnants
and magnetar ejecta~\citep{Le15, Ko07, Ku15}. Thus, if future observations
establish this distance scale for the FRB population, it should be 
possible to better constrain the model of absorption and the progenitor.

During the course of this work, we observed an interesting trend in the
FRB flux as a function of redshift for observations in the $<1$~GHz band
where models C and F predict an increase in flux density as a function
of redshift (see, e.g., the left panel of Fig.~\ref{fig:spec}). This 
behaviour is due to the Doppler shifting of a spectrum with a turn-over
in its rest frame, which is a natural feature of the free-free absorption
models. For sources at higher redshifts, we sample a different region 
in the spectrum of the source (see the right panel of Fig.~\ref{fig:spec}). 
If the spectrum has a turnover,
the peak flux increases as we sample the rising edge of the
spectrum. At higher redshifts, the frequency band passes over the
turnover resulting in a decrease in the peak flux as expected. 
As discussed in \S \ref{sec:caveats}, we have not included the
potential increase in the FRB rate with redshift that is predicted in
cosmological models invoking star formation~\citep{Ma14}. If these
models prove to be relevant in future, the aforementioned effect will
be even stronger than seen in Fig.~\ref{fig:spec}.

The constraints given in Table~\ref{tab:params} can tell us
  about the nature of the FRB progenitors. The observed and predicted
  spectral indices suggest that FRB spectral indices are different
  from pulsar spectral indices which have a mean of
  -1.4~\citep{Ba14}. Observations have suggested that at least some
  FRB spectral indices are positive \citep{Sp16}. Assuming a
  synchrotron source, the spectral index and the flux together can
  give us order of magnitude estimates about the magnetic field and
  effective electron temperature of the source~\citep[see for
    e.g][]{Co16c}. For example, if FRBs truly have a positive spectral
  index at frequencies of 1~GHz, the results favour a compact source
  with large magnetic field that is perpendicular to the line of sight
  (e.g., as seen in magnetar bursts) since the frequency at which the
  source becomes optically thick is proportional to the magnitude of
  the magnetic field while a negative spectral index would suggest
  other synchrotron sources (e.g., giant pulses from neutron stars).
  A large sample size of these sources expected from CHIME and HIRAX
  will definitely help to alleviate the problem.

In summary, we have carried out a detailed analysis of possible FRB
source populations and the expected yield from ongoing and future
radio surveys below 1~GHz, based on results from the previous
surveys. The previous results help in constraining the spectral index
of the burst although no inference on the emission model can be drawn
currently.  Even with the most stringent model, in which spectral
turnovers are expected in the observing band, CHIME is expected to see
FRBs very frequently. Similar results are expected to be seen by
HIRAX. The yields of CHIME, HIRAX and UTMOST will undoubtedly lead to
a large sample that will provide great insights into the nature of and
emission mechanism of these enigmatic sources.

\section*{Acknowledgments}

We thank Jayanth Chennamangalam for providing code to make plots for
Fig.~\ref{fig:N_z}, and Joeri van Leeuwen for pointing out the flux degradation effect
shown in Figs.~2, 3, 4 and 5.
We acknowledge the assistance of Kevin Bandura who provided
useful information necessary to carry out the predictions for CHIME and
HIRAX. We thank Victoria Kaspi, Manisha Caleb
and Pragya Chawla for useful discussions.

\bibliography{Ref}{}

\bibliographystyle{mn2e}
\end{document}